\documentstyle[epsfig,aps,floats,preprint]{revtex}

\newcommand{\beq}{\begin{equation}}
\newcommand{\eeq}{\end{equation}}
\newcommand{\bea}{\begin{eqnarray}}
\newcommand{\eea}{\end{eqnarray}}

\def\laq{~\raise 0.4ex\hbox{$<$}\kern -0.8em\lower 0.62
ex\hbox{$\sim$}~}
\def\gaq{~\raise 0.4ex\hbox{$>$}\kern -0.7em\lower 0.62
ex\hbox{$\sim$}~}

\def \pa {\partial}
\def \ra {\rightarrow}

\def \Da {\Delta}
\def \b {\beta}
\def \a {\alpha}

\def \ga {\gamma}
\def \sg {\sigma}
\def \da {\delta}

\def \om {\omega}

\def \ti {\tilde}

\begin{document}
\par
\begingroup
%\twocolumn[%

\begin{flushright}
BA-TH/99-363\\
September 1999\\
gr-qc/9910019\\
\end{flushright}
\vskip 1true cm

\vspace{12mm}
{\large\bf\centering\ignorespaces
 On the response of gravitational antennas\\ to dilatonic waves
\vskip2.5pt}

\bigskip
{\dimen0=-\prevdepth \advance\dimen0 by23pt
\nointerlineskip \rm\centering
\vrule height\dimen0 width0pt\relax\ignorespaces
M. Gasperini
\par}

{\small\it\centering\ignorespaces
Dipartimento di Fisica, Universit\`a di Bari, \\
Via G. Amendola 173, 70126 Bari, Italy \\
and \\Istituto Nazionale di Fisica Nucleare, Sezione di Bari,
Bari, Italy \\
\par}

\par
\bgroup
\leftskip=0.10753\textwidth \rightskip\leftskip
\dimen0=-\prevdepth \advance\dimen0 by17.5pt \nointerlineskip
\small\vrule width 0pt height\dimen0 \relax

\begin{abstract}
It is pointed out that the coupling of macroscopic test masses to the
gravi-dilaton background of string theory is non geodesic, in general,
and cannot be parametrized by a Brans-Dicke model of scalar-tensor
gravity. The response of  gravitational antennas to dilatonic waves
should be analyzed through a generalized equation of geodesic
deviation, taking into account the possible direct coupling of the
background to the (composition-dependent) dilatonic charge of the
antenna.
\end{abstract}

\par\egroup
%\vskip2pc]
\thispagestyle{plain}
\endgroup


%\section {Introduction}
%\label{I}

A number of papers has recently explored the interesting possibility
of detecting scalar waves, with coupling strength to matter of
gravitational order, exploiting both resonant mass \cite{1} -\cite{3} and
interferometric \cite{4,5} gravity wave detectors. Indeed,
gravitational antennas conceived to respond to the tensor part of
metric fluctuations can also respond to the scalar oscillations of the
metric background, induced by the coupling to some fundamental
scalar component of the gravitational multiplet. 

Such papers are basically motivated by the possible emission of scalar
waves in astrophysical process, like the spherically symmetric collapse
of a star, or of a cloud of dust \cite{2,6}. Another motivation is the
possible presence or a relic cosmic background of (light enough) scalar
particles. Such particles could be  produced copiously in the early
Universe (see for instance \cite{7,8}), and could survive until
today in the form of a stochastic background of scalar waves,  
representing a possible significant fraction of the present large scale
density of dark matter. Given the extreme weakness of their 
couplings,  gravitational antennas are probably (at present) the only
plausible candidates for their {\sl direct} detection \cite{9}. 

In all the quoted papers \cite{1} - \cite{5}, the analysis of the possible
response of  the detector was performed assuming a {\em geodesic}
coupling of the test masses (representing the detector) to the {\em
scalar component} of the metric fluctuations. The standard equation of
geodesic deviation was subsequently applied to estimate the detector
sensitivity. This is certainly justified when the gravitational
scalar-tensor interactions are parametrized by a strictly
``Brans-Dicke type" model of gravity, as always assumed  in \cite{1} -
\cite{5}. In that case, a ``Jordan frame" exists in which there are no
couplings to the scalar field in the matter part of the action, and the
scalar interactions are totally absorbed in the rescaled metric. 

This is not the most general case, however, and  one of the purposes of
this paper is to point out that this {\em is not} the case, in particular, 
for the coupling of macroscopic test masses to the scalar dilaton field
appearing in unified theories of strings and superstrings 
\cite{10}.  In that context, in fact, ordinary macroscopic masses have a
scalar ``dilatonic charge", which is in general {\em non-universal}
\cite{11}, depending on the internal, nuclear composition of the given
body. It is thus in general  impossible to define a universal Brans-Dicke
frame in which the scalar interactions are absorbed in the rescaled
metric, for all test  bodies. 

As a consequence, the equations of motion of test masses, and the
response of gravitational antennas,  should included in general a direct,
non-geodesic coupling of the dilaton charge of the test body to the
gradients of the external dilaton field. Such couplings will eventually
provide an additional, explicit dilatonic contribution to the equation of
geodesic deviation. For string models, this is true even in the so-called
``String frame", in which the gravi-dilaton action takes the form of an
effective Brans-Dicke action. 

Let me start recalling some elementary notion concerning the motion
of a test body, described by the Lagrangian $L_m$, coupled to a 
scalar-tensor bakground characterized by  the Brans-Dicke
parameter $\om$: 
\beq
S= \int d^4x \sqrt{-g} ~e^{-\phi} \left[-R+\om \left(\nabla \phi
\right)^2\right]+\int d^4x \sqrt{-g}L_m 
\label{1}
\eeq
(metric conventions: $+---$). The variation with respect to the metric 
and to the dilaton $\phi$ gives, respectively, the field equations (in
units $16\pi G=1$):
\bea
&& 
G_{\mu\nu} + \nabla_\mu\nabla_\nu\phi -(\om+1) 
\nabla_\mu\phi\nabla_\nu\phi -g_{\mu\nu} \nabla^2 \phi
+\left({\om\over 2} +1\right)g_{\mu\nu} \left(\nabla \phi\right)^2
={1\over 2} e^\phi T_{\mu\nu}, 
\label{2}\\
&&
R +\om \left(\nabla \phi\right)^2 - 2 \om \nabla^2 \phi+e^\phi \sg =0,
\label{3}
\eea
where $G_{\mu\nu}=R_{\mu\nu} -g_{\mu\nu}R/2$ is the Einstein tensor,
and 
\beq
T_{\mu\nu}= {2\over \sqrt{-g}} {\da (\sqrt{-g} L_m) \over \da
g^{\mu\nu}}, ~~~~~~~~~
\sg= {1\over \sqrt{-g}} {\da (\sqrt{-g} L_m) \over \da\phi},
\label{4}
\eeq
are, respectively, the energy momentum and dilatonic charge density
of the test body. For $\sg \equiv 0$, the combination of the above
equations, and the use of the Bianchi and Riemann identities,
\beq
\nabla^\nu G_{\mu\nu}=0, ~~~~~~~~~
\left[\nabla^\a \nabla_\mu \phi,  \nabla_\mu \nabla^\a \phi\right]
\nabla_\a \phi = R_\mu \,^\nu \nabla_\nu \phi,
\label{5}
\eeq
leads immediately, for {\em any} $\om$, to the covariant conservation
of the stress tensor, $\nabla^\nu T_{\mu\nu}=0$, and to the consequent
geodesic motion of the test body. 

If $\sg \not= 0$, however, this result is no longer valid. Let me discuss,
for simplicity, the physically interesting case $\om =-1$, corresponding
to the lowest-order gravi-dilaton effective action of string theory, in
the String frame. By applying the covariant differential
operator to the above equations we get 
\beq
\nabla^\nu T_{\mu\nu}+\sg \nabla_\mu \phi=0,
\label{6}
\eeq
which can also be written as
\beq
\pa_\nu \left(\sqrt{-g} T^{\mu\nu}\right) + \sqrt{-g}
\Gamma_{\a\nu}\,^\mu T^{\a\nu} + \sqrt{-g} \sg \nabla^\mu \phi=0
\label{7}
\eeq
(for $\om \not= -1$ there are additional, $\om$-dependent, dilaton
terms). 

In order to check explicitly that the motion is not geodesic we can 
apply the so-called multipole expansion \cite{12}, assuming that the
gravitational and dilatonic charges of the test body are nonzero only
inside a thin ``world-tube", centered around the world line
$x^\mu(\tau)$ of the center of mass. Inside the world-tube, we expand
the external fields  $\{ \Gamma, \nabla \phi\}$ around the position
$x^\mu$ of the center of mass: 
\bea
&&
\Gamma_{\a\nu}\,^\mu(x') = \Gamma_{\a\nu}\,^\mu(x)+
(x'-x)^\b \pa_\b \Gamma_{\a\nu}\,^\mu(x)+ ...
\nonumber\\
&&
\nabla^\mu \phi(x') = \nabla^\mu \phi(x)+
(x'-x)^\b \pa_\b \nabla^\mu \phi(x)+ ...
\label{8}
\eea
By integrating eq. (\ref{7}) on the spacelike hypersurface $x^0=$ const,
and neglecting internal momenta in the point-particle (or pole-particle
\cite{12}) approximation, we get 
\beq
{d\over dx^0} \int d^3 x' \sqrt{-g} T^{\mu 0}(x')+ 
\Gamma_{\a\nu}\,^\mu(x)\int d^3 x' \sqrt{-g} T^{\mu
\nu}(x')+\nabla^\mu \phi (x) \int d^3 x' \sqrt{-g} \sg(x'),  
\label{9}
\eeq
where the coordinates  $x'$ range inside the three-dimensional
space-like section of the world tube (the spatial divergence, $\pa_i 
(\sqrt{-g} T^{\mu i})$, has been eliminated  through the Gauss theorem). 

We recall now that, in the point-like limit, the generally covariant
stress tensor for a particle of mass $m$, and world-line $x^\mu(\tau)$,
is given by \cite{13}
\beq
T^{\mu\nu}(x')= {p^\mu p^\nu \over \sqrt{-g}p^0}\da^{(3)}
\left(x'-x(\tau)\right),
\label{10}
\eeq
where $p^\mu= mu^\mu=m dx^\mu/d\tau$. We can define, in the same
way, the dilaton charge density in terms of the dimensionless, relative
strength $q$ of scalar to tensor forces for the given test body (i.e. the
scalar charge per unit of gravitational mass) as: 
\beq
\sg(x')= q {m^2\over \sqrt{-g}p^0}\da^{(3)}
\left(x'-x(\tau)\right) 
\label{11}
\eeq
(the net dilaton charge $q$ may be different for different test bodies,
see below). By integrating eq. (\ref{9}) in the limit in which
the radius of the world tube shrinks to zero, $x' \ra x$, and multiplying
by $m^{-2} p^0=m^{-1} dx^0/d\tau$, we get finally
\beq
{d u^\mu\over d\tau} + \Gamma_{\a\nu}\,^\mu u^\a u^\nu + q 
\nabla^\mu \phi=0.
\label{12}
\eeq
The motion of the given test body is clearly non-geodesic in a
non-trivial gravi-dilaton background, $\nabla \phi \not= 0$ (by the
way, this equation implies that the dilaton has to be a short-range
field if $ q\gaq 1$, to avoid contradictions with tests of the
equivalence principle \cite{11,12a}; but I will come back on this point
later). 

It is now an easy task to compute the dilaton corrections,
induced by the charge $q$, to the standard equation of geodesic
deviation \cite{13} used to analyze the response of gravitational
antennas. We consider two infinitesimally close world-lines,
$x^\mu(\tau)$ and $x^{\prime\mu}(\tau)$, satisfying eq. (\ref{12}), and
differing by the spacelike separation vector $\eta^\mu$, namely 
$x^{\prime \mu}(\tau)= x^\mu(\tau)+ \eta^\mu(\tau)$. In the
equation of motion for $x^{\prime \mu}(\tau)$ we expand the
external fields as in (\ref{8}), and using the motion of $x^\mu$ we
obtain an expression for the  acceleration of the separation vector, 
$d^2 \eta^\mu/d\tau^2$. Shifting to covariant derivatives,
$D/D\tau$, the final expression can be written in compact form as
\beq
{D^2 \eta^\mu \over D\tau^2} + R_{\b\a\nu}\,^\mu u^\a u^\b \eta^\b+
q\eta^\b \nabla_\b\nabla^\mu \phi =0.
\label{13}
\eeq
This gives the covariant, relative acceleration of two neighbouring
world-lines, for a test body coupled with a charge $q$ to a non-trivial
gravi-dilaton background. 

We come now to the important point already emphasized at the
beginning of this paper. Why the motion of realistic macroscopic test
masses, in the gravi-dilaton background of string theory, may be
expected to be non-geodesic even in the String frame? In other words,
why macroscopic test bodies, in a string theory context, may have 
composition-dependent dilatonic charges and, consequently, are not
adequately described by a pure Brans-Dicke type model of
gravity? 

To answer this question let me recall that the fundamental fields
building up ordinary macroscopic matter, including all loops in the
string effective action, are in general coupled non-minimally and
non-universally to the dilaton. The weak coupling limit of the dilaton
charge $q$ has been carefully estimated in the Einstein frame \cite{7,11}
through a canonical rescaling of fields and masses, and found to
depend on two computable (in principle), non-universal loop functions.
General arguments  then suggest  $q\gaq 1$ for hadronic matter
(typically, $q \sim 44$ for nucleons), while $q\sim 1$ for leptons 
\cite{11,12b}.  
I will not repeat the computations of \cite{11,7}, but I will
show here that in the String frame the effective charges are also
non-vanishing, and typically of the same order as in the Einstein frame. 

To this aim, let me consider a scalar field model matter, $\psi_i$, whose
gravi-dilaton interactions, including all possible loop corrections, are
described by the effective action, in $d+1$ dimensions:
\beq
S= \int d^{d+1}x \sqrt{-g} \left[ -Z_R (\phi) R -Z_\phi (\phi)\left(\nabla
\phi\right)^2 -V(\phi) + {1\over 2} Z_k^i (\phi)\left(\nabla
\psi_i\right)^2 + Z_m^i (\phi)\psi_i^2 \right].
\label{14}
\eeq
Here $Z^i$ represent the dilaton coupling function of the field $\psi_i$,
computed in the String frame, where $g_{\mu\nu}$ is the metric of the
conformal sigma-model describing the motion of fundamental strings
in the given background. It is always possible, however, to restore the
canonical form of all the kinetic terms in the action by introducing a
set of rescaled fields $\{\ti g_{\mu\nu}, \ti \phi, \ti \psi\}$, defined in
terms of the old fields and of the coupling functions $Z(\phi)$ (see for
instance \cite{7}). The action becomes
\beq
S= \int d^{d+1}x \sqrt{-\ti g} \left[-\ti R +{1\over 2} \left(\ti \nabla
\ti \phi_i\right)^2 - \ti V(\ti \phi)+ {1\over 2} \left(\ti \nabla
\ti \psi_i\right)^2+ L(\ti \phi, \ti \psi_i)\right], 
\label{15}
\eeq
where $\ti \nabla$ is the covariant derivative for the Einstein metric
$\ti g_{\mu\nu}$, and
\beq
 L(\ti \phi, \ti \psi_i)\equiv  {1\over 2} \ti \mu_i^2(\ti \phi)\ti
\psi_i^2, ~~~~~~~~~~~
\ti \mu_i^2(\ti \phi)= Z_m^i \left[Z_k^i\right]^{-1} 
\left[Z_R\right]^{2/(1-d)}, 
\label{16}
\eeq
is the canonical matter-dilaton interaction Lagrangian. Its low-energy
expansion around the value of $\ti\phi$ which extremizes the dilaton
potential (and which can always be assumed to coincide with
$\ti\phi=0$, after a trivial shift), 
\beq
 L(\ti \phi, \ti \psi_i)= {1\over 2} m_i^2 \ti \psi_i^2 +
{1\over 2} \ti g_i  \ti \phi \ti \psi_i^2+ ...
\label{17}
\eeq
defines the effective masses and dilaton couplings as 
\bea
&&
m_i^2= \left[\ti \mu_i^2(\ti \phi)\right]_{\ti\phi=0},
\nonumber\\
&&
\ti g_i = \left[{\pa\over \pa \ti\phi }\ti \mu_i^2(\ti
\phi)\right]_{\ti\phi=0}= 
m_i^2\left[{\pa\over \pa \ti\phi }\ln\ti \mu_i^2(\ti
\phi)\right]_{\ti\phi=0}. 
\label{18}
\eea
In the weak coupling limit $Z_R=Z_\phi=e^{-\phi}$. 
For $d=3$, in addition, one finds \cite{7} $\phi=\ti\phi$, and the
relative coupling strength of scalar to tensor forces (i.e. the dilaton
charge in units of the ``gravitational charge" $\sqrt{16 \pi G} m_i$)
becomes \cite{11,7}, according to eq. (\ref{16}):
\beq
\ti q_i ={\ti g_i\over m_i^2}=1+ \left[{\pa\over \pa \phi }\ln \left(
Z^i_m \over Z^i_k \right)\right]_{\phi=0}. 
\label{19}
\eeq

In the String frame, on the other hand, the canonical matter field
$\widehat \psi_i=[Z_k^i]^{1/2}\psi_i$ is defined with respect to the
unrescaled metric $g_{\mu\nu}$, and the matter-dilaton interaction
Lagrangian becomes, at low energy (see eq. (\ref{14})): 
\beq
 L( \phi, \widehat\psi_i)=  {1\over 2}\mu_i^2( \phi)
\widehat\psi_i^2, ~~~~~~~~~~~
\mu_i^2= Z_m^i \left[Z_k^i\right]^{-1} . 
\label{20}
\eeq
In the weak coupling limit, this gives a dimensionless dilaton charge
\beq
 q_i =\left[{\pa\over \pa \phi }\ln \left(
Z^i_m \over Z^i_k \right)\right]_{\phi=0}= \ti q_i-1. 
\label{21}
\eeq
Thus, unless $\ti q_i$ is fine-tuned to unity, the dilaton charge is
non-universal and non-vanishing both in the String and Einstein frame. 

For a macrosopic body the total charge $q$, in the weak field limit, is
obtained by summing over all the components, $q=\sum_i m_i
q_i/\sum_i m_i$. Suppose that the body, of mass $M$, is composed of
$B$ baryons with mass and charge $m_b, q_b$, and $Z$ electrons 
with mass and charge $m_e, q_e$. For $Z\sim B$, $m_e \ll
m_b$, we obtain: 
\beq
q \simeq Bm_b q_b/M= \left(B/\mu\right)q_b,
\eeq
where $\mu=M/m_b$ is the mass of the body in units of baryonic
masses. Since $B/\mu \sim 1$, the total dilaton charge is controlled by
the dilaton coupling to baryons, $q_b$. If, as suggested in 
\cite{11,12b},   
$\ti q_b  \gg 1$, then the dilaton charge of macroscopic bodies is
of the same order of magnitude both in the String and Einstein frame,
and is composition-dependent (as $B/\mu$ depends on the internal
nuclear structure), with variations, across different types of ordinary
matter, which are typically of order 
\beq
{\Da q\over q} \simeq \Da \left(B\over \mu\right) \sim 10^{-3}.
\eeq

Let me come back, finally, to eq. (\ref{13}), which applies to all models
of gravity (not only string theory) in which macroscopic test masses
are coupled non-geodesically, with dilatonic charges $q$, to an
external scalar-tensor background. Such equation (never considered
previously in the gravity-wave literature, to the best of my
knowledge), should be taken as the starting point for a general
analysis of the response of a detector to scalar oscillations. A detailed
computation of the sensitivity to dilatonic waves, based on such
equation, is outside the purpose of this paper, and is demanded to
further studies and to the work of research group with experience in
the analysis of gravitational antennas. However, let me attempt here a
naive, qualitative estimate, just to have a first  indication of the
possible differences induced by the direct coupling to the dilatonic
charge of the antenna. 

I will follow the standard analysis presented in \cite{14}. For small,
non-relativistic oscillations of two test masses, with rest separation
$L^\mu$, we can put $\eta^\mu= L^\mu + \xi^\mu$, and from eq.
(\ref{13}) we obtain, to first order,
\beq
\ddot \xi^i + R_{koo}\,^iL^k +q L^k \pa_k \pa^i \phi =0.
\label{22}
\eeq
We include restoring and damping mechanical forces, corresponding to
a proper oscillation frequency $\om_0$, and to a friction coefficient
$\ga$:
\beq
\ddot \xi^i +\ga \dot \xi^i +\om_o^2\xi^i + R_{koo}\,^iL^k +q L^k \pa_k
\pa^i \phi =0. 
\label{23}
\eeq
Consider the response to a monocromatic scalar wave $\phi=\phi_0
\exp(ikx -i\om t)$, propagating along the oscillator direction. As we
are interested in the direct coupling to the dilatonic charge, we neglect
here the gravitational coupling to the metric, and we get the
steady-state solution
\beq
\xi(t)= {qLk^2\over \om^2-\om_0^2+i\ga \om}\phi_0e^{-i\om t}
\label{24}
\eeq
(the spatial dependence of the wave has been neglected, assuming
$|kx|\ll1$ throughout the oscillator). For a detector with two masses
$M$, vibrating with the above amplitude, we can define a vibration
energy $M\dot\xi^2$, and a corresponding cross section $\sg$ (the
energy dissipation rate per unit of incoming flux) as \cite{14}
\beq
\sg_\phi ={M\dot\xi^2 \ga\over \om^2 \phi_0^2}.
\label{25}
\eeq

Up to this point, the response of the detector to scalar radiation is
similar to the case of tensor radiation, with two important differences,
however: the additional presence of the dilaton charge, which multiplies
the incident amplitude, and the fact that the {\em time derivatives} of
the metric oscillations, $\ddot h$ (contained in the Riemann tensor), are
replaced by the {\em spatial gradients} of the scalar wave, $\nabla^2
\phi$ (coupled to the dilaton charge). As a consequence, the response
(\ref{24}) of the detector turns out to be proportional to $k^2$ instead
of $\om^2$. We have now two possibilities.

$1)$ The scalar waves are massless. In this case $k=\om$, and the
reponse is greatly enhanced for waves of resonant frequency
$\om=\om_0$. The cross section, at the resonance,
\beq
\sg_\phi ={Mq^2 L^2\om_0^2\over \ga}
\label{26}
\eeq
is the same as the graviton cross section, multiplied however by the
dilaton charge $q^2$. For the string theory dilaton, unfortunately, the
charge in this case gives a very strong suppression factor, because a
massless dilaton is incompatible with the 
present tests of the equivalence 
principle unless some form of universality is assumed in the string
loop corrections \cite{15}, thus evading the conclusion of \cite{11}, and
leading to a very small dilatonic coupling to matter, $q^2 \ll1$. 
This suppression should be valid for any long-range scalar field 
with composition-dependent couplings. Assuming that the detectors are
sensitive enough, however, it should be possible, in this case, to 
discriminate scalar from tensor signals by comparing the response of
different antennas. If, on the contrary, the long range scalar field is 
universally coupled to matter, then a Brans-Dicke model of gravity 
is appropriate, and the analysis performed in \cite{1} - 
\cite{5} can be applied.

$2)$ The scalar waves are massive. In this case the response of the
detector is model-dependent. For the string theory dilaton, in
particular, the coupling may be large, $q>1$, but then the dilaton mass
has to be large enough \cite{11,12a,7}, i.e. $m \gaq 10^{-4}$ eV $\sim
10^{2}$ GHz, to keep the dilaton corrections to macroscopic gravity
below the threshold of present experimental observations \cite{16}. 
Since $\om > m$, it seems impossible in this case to match the
resonance condition of present gravity wave detectors, with $\om_0
\sim 10^2-10^3$ Hz. We have always $\om \gg \om_0$, $ \om \gg
\ga$,  and the cross section becomes 
\beq
\sg_\phi =Mq^2 L^2\ga \left(k\over \om\right)^4.
\label{27}
\eeq
With respect to the resonant graviton cross section, 
$\sg_h =M L^2\om_0^2/\ga$, the response of the detector to massive
dilatons, of momentum $k$ and energy $\om$, is thus suppressed by the
ratio 
\beq
{\sg_\phi\over \sg_h} = {q^2\over Q^2} \left(k\over \om\right)^4 \ll 1,
\label{28}
\eeq
where the factor $Q=\om_0/\ga$ is in general very large for resonant
detectors \cite{14}. 

It should be noted, as a final remark, that in a scalar-tensor model of
gravity the oscillations of the scalar background induce oscillations in
the scalar sector of the fluctations of the metric. The scalar
oscillations of the metric contribute to the total Riemann tensor, and
are thus gravitationally coupled to the detector, through the standard
equation of geodesic deviation. This is the effect considered in previous
papers \cite{1} - \cite{5}, which is contained in the Riemann part of eq.
(\ref{23}), and which is to be added to eq. (\ref{24}) when computing
the full response of the detector. 

In some models, the direct coupling to the dilaton charge of the
antenna could  represent the dominant factor, controlling the response
to scalar radiation. Even in that case, however, the results about the
possible detection of (massive or massless) dilatons predicted in a 
string theory context seem to remain pessimistic, in agreement with the
conclusions of \cite{5}. Nevertheless, 
it may be important to note that for 
dilatons, unlike for gravitons, the direct coupling of the scalar wave to
the dilatonic charge of the antenna induces a response proportional to
the momentum squared (instead of the energy squared) of the incident
particles. Can this help for the experimental detection of massive scalar
waves?

\acknowledgments
I wish to thank  Maura Brunetti, Michele Maggiore and 
Gabriele Veneziano for useful discussions and 
comments.

\end{document}